\documentstyle[prl,aps,multicol,epsf]{revtex}
\newcommand{\beq}{\begin{equation}}
\newcommand{\eeq}{\end{equation}}

\begin{document}

\title{Influence of collective effects and the d-CDW on electronic Raman
scattering in high-T$_c$ superconductors}

\author{{R. Zeyher}$^a$ and {A. Greco}$^{a,b}$} 

\address{$^a$Max-Planck-Institut\  f\"ur\
Festk\"orperforschung,\\ Heisenbergstr.1, 70569 Stuttgart, Germany \\
$^b$Permanent address: Departamento de F\'{\i}sica, Facultad de
Ciencias Exactas e Ingenier\'{\i}a and \\
IFIR(UNR-CONICET), Av. Pellegrini 250, 
2000-Rosario, Argentina}

\date{\today}


\maketitle

\begin{abstract} 
Electronic Raman scattering in high-T$_c$ superconductors is studied
within the $t-J$ model. It is shown that the
$A_{1g}$ and $B_{1g}$ spectra are dominated by amplitude
fluctuations of the superconducting and the d-wave CDW order parameters,
respectively. The $B_{2g}$ spectrum contains no collective effects 
and its broad peak reflects vaguely the doping dependence of $T_c$, similarly
to the pronounced peak in the $A_{1g}$ spectrum. 
The agreement of our theory with the 
experiment supports the picture of two different, competing order 
parameters in the underdoped regime of high-T$_c$ superconductors.

\end{abstract}

\begin{multicols}{2}


Electronic Raman scattering in superconductors probes charge excitations
across the superconducting gap and thus provides
information on the magnitude and the anisotropy of the 
gap\cite{Kendziora,Strohm,Opel,Venturini}.  
Commonly used weak-coupling theories for the interpretation of the data are,
however, not able to
account for the experimental spectra in the cuprates,
especially, as a function of doping. Such theories yield values for the
d-wave gap which 
have no simple relation to $T_c$ 
and which are substantially smaller than 
those measured in ARPES experiments\cite{Norman}. In our view, these problems 
arise because electronic correlation effects are not adequately taken into 
account. Below we consider a strong-coupling model
which allows to consider interaction effects between excited quasi-particles
as well as the appearance of a d-CDW at lower dopings 
in a systematic way.

  Our calculations of Raman response functions are based on the 
widely accepted t-J model
where the two spin components have been generalized to N components
and the leading  diagrams at large N are taken into account.
As discussed in detail in Ref.\cite{Cappelluti} the  phase diagram 
in this
limit is largely determined by the onset of a flux phase with d-wave symmetry,
often also called d-CDW\cite{Chakravarty}, at a doping $\delta = \delta_0$. 
The corresponding order parameter has the form
$\Phi({\bf k}) = {{-i}/{2N_c}}\sum_{{\bf q}\sigma}J({\bf k}-{\bf q})
\langle {\tilde c}^\dagger_{{\bf q}\sigma}\tilde{c}_{{\bf q + Q}\sigma}
\rangle$. $J$ is the Heisenberg coupling, $\tilde{c}^\dagger,\tilde{c}$
are creation and annihilation operators for electrons under the
constraint that double occupancies of lattice sites are excluded, $N_c$
is the number of primitive cells, $\langle ...\rangle$ 
denotes an expectation value, and $\bf Q$ is the wave vector of the d-CDW.
Furthermore, there exists for all dopings an instability towards d-wave 
superconductivity\cite{Zeyher}. Keeping only the instantaneous term in the 
effective interaction, the order parameter is 
$\Delta({\bf k}) = {1/{2N_c}} \sum_{\bf q} (J({\bf k}-{\bf q}) -
V_C({\bf k}-{\bf q})) \langle {\tilde c}_{{\bf q}\uparrow} 
{\tilde c}_{-{\bf q}\downarrow}
\rangle$.
As shown in Ref.\cite{Cappelluti} it is in general necessary to include 
the Coulomb potential V$_C$ in order to stabilize the d-CDW with respect 
to phase separation. The resulting superconducting transition temperature 
$T_c$ decreases with increasing doping for
$\delta > \delta_0$. For $\delta < \delta_0$ the two order parameters compete
with each other leading to a strongly decreasing $T_c$ with decreasing 
doping. As a result
optimal doping is essentially determined by $\delta_0$. In the presence
of the two order parameters the operators
$({\tilde c}^\dagger_{{\bf k},\uparrow},\tilde{c}_{-{\bf k},\downarrow},
{\tilde c}^\dagger_{{\bf k+Q},\uparrow},\tilde{c}_{{\bf -k-Q},\downarrow})$
are coupled leading to the following Green's function matrix\cite{Cappelluti}
\begin{equation}
G^{-1}(z,{\bf k}) = \left( 
\begin{array}{c c c c}  
z-\epsilon({\bf k}) & -\Delta({\bf k})  & -i\Phi({\bf k})                      &  0                 \nonumber\\
-\Delta({\bf k})    &z+\epsilon({\bf k})
&   0               &i\Phi({\bf \bar{k}})      \\
i\Phi({\bf k})      &   0
&z-\epsilon({\bf\bar{k}})& -\Delta({\bf \bar{k}})  \nonumber\\
          0         &-i\Phi({\bf \bar{k}})
&-\Delta({\bf\bar{k}})&z+\epsilon({\bf \bar{k}})  

\end{array} \right)
\label{matrix}
\end{equation}
$\epsilon({\bf k})$ is the one-particle energy, 
$\epsilon({\bf k}) = -(\delta t +\alpha J)(cos(k_x)+cos(k_y))
-2t'\delta cos(k_x)cos(k_y)  -\mu$,
with $\alpha = 1/N_c \sum_{\bf q} cos({q_x})f(\epsilon({\bf q}))$.
$f$ is the Fermi function, $\delta$ the doping away fom half-filling,
$\mu$ a renormalized chemical potential,
$t$ and $t'$ are nearest and second-nearest
neighbor hopping amplitudes, z a complex frequency, and $\bf {\bar{k}}
= {\bf k-Q}$.

Expressing the expectation values in the order parameters by $G$ and
using Eq.(\ref{matrix}) one obtains two coupled equations for the order 
parameters.
Detailed considerations\cite{Cappelluti} show that for dopings in the 
neighborhood of
the optimal doping $\delta_0$ the most stable order parameters have
d-wave symmetry, $\Phi({\bf k})=\Phi \gamma({\bf k}),
\Delta({\bf k})=\Delta \gamma({\bf k})$, with 
$\gamma({\bf k}) =(cos(k_x)-cos(k_y))/2$. Fig. 1 shows the doping dependence
of $\Phi$ and $\Delta$ at zero temperature, calculated for $t'/t=-0.35$
and $J/t=0.3$. The energy unit is $t$.
A repulsive nearest-neighbor Coulomb interaction was also included
with $V_C/t=0.06$. In the overdoped region $\delta > \delta_0$ $\Phi$ is zero
and $\Delta$ increases monotonically with decreasing $\delta$. After the onset
of the flux phase at $\delta_0$ $\Phi$ is first suppressed by superconductivity
and then, with decreasing doping, increases steeply and suppresses now,
in addition to V, the superconducting order parameter. In our calculation we
assumed the value $(\pi,\pi)$ for the wave vector $\bf Q$. Fig. 1 shows that 
superconductivity coexists in the underdoped region with the d-CDW and that 
the competition between the two order parameters leads to the rapid decay
of $T_c$ towards small dopings. 
\begin{figure}[h]
\centerline{
      \epsfysize=6cm
      \epsfxsize=7cm
      \epsffile{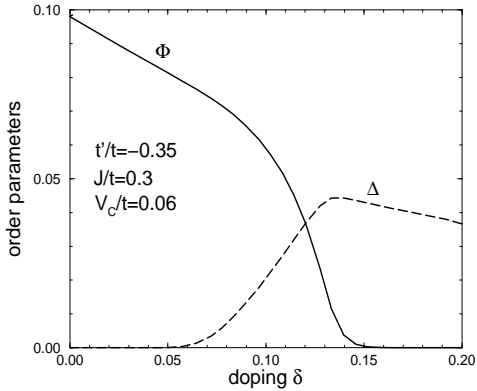}}
\label{fig1}
\caption 
{Order parameters $\Phi$ and $\Delta$
as a function of doping \\
in units of $t$ at $T=0$.}
\end{figure}
\noindent
The scattering intensity of electronic Raman scattering 
is proportional to
$-(1+n(\omega)) Im \chi_\alpha (\omega+i\eta)$, where $n$ denotes the 
Bose factor,
$\omega$ the frequency, and $\chi_\alpha$ a response function of
the non-local density operator $\rho_\alpha({\bf k})=1/N_c
\sum_{{\bf k},\sigma}\gamma_\alpha({\bf k}) {\tilde c}^\dagger_{{\bf k},
\sigma}{\tilde c}_{{\bf k} \sigma}$. $\alpha=1,3,4$ stands for 
the representations $A_{1g},B_{1g},B_{2g}$ of the point 
group $D_{4h}$ of the square lattice, respectively. In the large N limit of
the t-J model we have $\gamma_1({\bf k})
= (\delta t +J \alpha)(cos k_x +cos k_y)/2 +2\delta t' cos k_x \cdot cosk_y$,
$\gamma_3({\bf k})= (\delta t +J\alpha)(cos k_x -cos k_y)/2$, 
$\gamma_4({\bf k}) = -2\delta t' sin k_x\cdot sink_y$. The response function
$\chi_\alpha$ is determined at large N by the sum over ladder diagrams 
shown in Fig. 2. 
\begin{figure}[h]
      \centerline{\hbox{
      \epsfysize=2cm
      \epsfxsize=4cm
      \epsffile{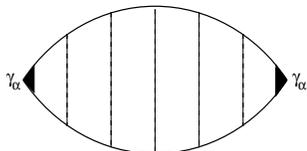}}}
\caption 
{Ladder diagrams for the Raman susceptibility; the \\
filled triangle is
the Raman vertex $\gamma_\alpha$, the solid and dashed \\
lines
represent the Green's function $G$ and a residual interaction,
respectively.}
\label{fig2}
\end{figure}

   In the case of the $B_{1g}$ and $B_{2g}$ spectra the residual interaction
is the Heisenberg interaction. Decomposing it into
irreducible basis functions of $D_{4h}$ it contains one $B_{1g}$ kernel 
but no contribution with $B_{2g}$ symmetry. This means that $\chi_4$
is given by the free susceptibility $\chi_4^{(0)}$ which
reduces to the standard expression 
if the d-CDW order parameter vanishes. 
We find the following analytic expressions
$\chi_3(z) = \chi^{(0)}_3(z)/(1+J'\chi^{(0)}_3(z))$,
$J'=J/(\delta t +J\alpha)^2$,
$\chi_4(z) = \chi^{(0)}_4(z)$, with
\noindent
\begin{eqnarray}
\chi^{(0)}_\alpha(z) = {2 \over N_c} \sum_{\bf k} 
\gamma^2_\alpha({\bf k}) \Bigl( \Pi_{11,11}({\bf k},z)
-\Pi_{12,21}({\bf k},z) \nonumber\\
+(-1)^\alpha [\Pi_{13,31}({\bf k},z)-\Pi_{14,41}({\bf k},z)]
\Big),
\label{chi0}
\end{eqnarray}
\begin{equation}
\Pi_{ij,kl}({\bf k},z) =  T\sum_n G_{ij}(i\omega_n+z,{\bf k})
G_{kl}(i\omega_n,{\bf k}),
\label{Pi}
\end{equation}
for $\alpha=3,4$.

Fig. 3 shows $B_{1g}$ and $B_{2g}$ spectra for
the parameters $t'/t=-0.35$, $J/t=0.3$, and $T=0$.
The upper diagram, corresponding
to $\delta = 0.178$, $\Delta=0.040$ and $\Phi = 0$, is typical for the 
overdoped region.
In the $B_{2g}$ spectra $\gamma_4$
heavily weights transitions near the diagonals in ${\bf k}$-space where
the gap vanishes. As a result the spectrum is broad and peaks
substantially below $2\Delta$. 
Neglecting vertex corrections $\chi_3^{(0)}$ is determined by 
free particle-hole
excitations mainly across the maximum of the d-wave gap. 
This leads to a well-pronounced peak at $2\Delta$ as shown by the dashed 
curve in the upper diagram of Fig.3. Including also vertex corrections
$\chi_3$ develops a bound state well within the d-wave gap which is
somewhat broadened by the finite density of states in the d-wave
gap. Practically all spectral weight is shifted from
the region around and above $2\Delta$ into the bound state which, in
anlogy to semiconductors, can be viewed as an exciton state\cite{Chubukov}.

The middle and lower diagrams in Fig. 3 decribe a slightly 
and a strongly underdoped case, respectively. In the extreme case of
$\Phi >> \Delta$ the density of quasi-particles shows a small gap due
to superconductivity embedded into a larger gap structure due to the flux
phase. The $B_{2g}$ spectra (dotted lines) are always very broad with a
maximum which is located roughly in the middle of the BCS-gap which decreases
with decreasing doping. In addition it contains a weak and
structureless background which extends over the whole gap region.
The dashed lines in Fig. 3
describe the $B_{1g}^{(0)}$ spectrum due to non-interacting particle-hole 
excitations.
It is rather insensitive to the BCS-part of the gap and consists of one
strong peak around the maximal total gap. Taking also vertex corrections 
into account practically the total spectral weight of the $B_{1g}^{(0)}$ 
curve is shifted into one peak which monotonically increases with
decreasing doping. It describes amplitude fluctuations 
of the d-CDW order parameter.  

The essential features of Fig.3 agree with experiments in the
cuprates, in particular, the increase of the frequency of the $B_{1g}$ peak 
with decreasing
doping and the nonmonotonic behavior of the $B_{2g}$ peak as a function of 
doping similar to that of $T_c$\cite{Opel}.  
Quantitative fits to experimental curves\cite{Devereaux} need, however, 
the inclusion of additional interactions, e.,g., with impurities, and are
therefore not attempted here. 
To what extent the predicted
large excitonic effect in the $B_{1g}$ spectrum is compatible with
experiment is presently unclear: The larger gap values deduced from ARPES
data in comparison to the Raman $B_{1g}$ peak in the
\begin{figure}[h]
      \centerline{\hbox{
      \epsfysize=10cm
      \epsfxsize=10cm
      \epsffile{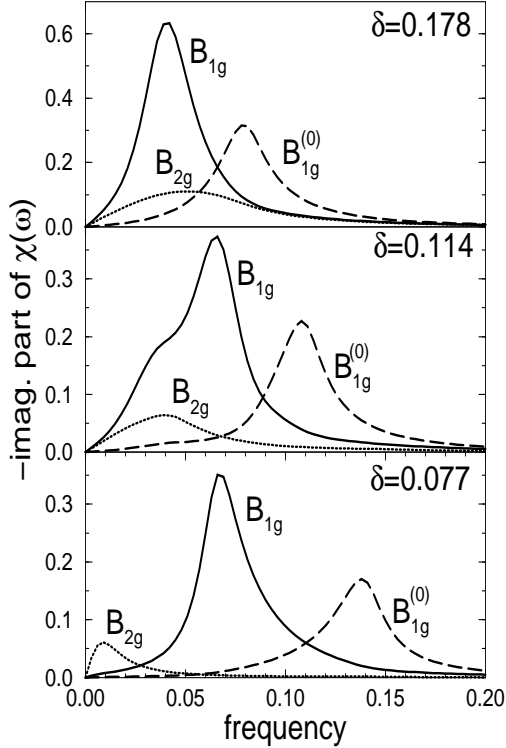}}}
\caption 
{Electronic Raman spectra of B-symmetry for three \\
different dopings
$\delta$, calculated for $t'/t=-0.35$, $J/t=0.3$, and \\
$T=0$.
The $B_{2g}$ spectrum has been multiplied by 10 in the \\
two upper 
and by 100 in the lower diagrams.}
\label{fig3}
\end{figure}
\noindent 
underdoped regime and the small difference between $B_{1g}$ and $B_{2g}$
peaks on the overdoped side are in favor of excitonic 
effects in the $B_{1g}$ channel. ARPES gives in slightly overdoped
$Bi2212$ $2\Delta \sim 70-75 meV$\cite{Norman} whereas the $B_{1g}$ peak 
lies near $59 - 63 meV$\cite{Kendziora,Venturini}, for the same doping. 
ARPES yields in untwinnned $YBa_2CuO_7$ gaps of 58 and 88 meV at the $X$ and 
$Y$-points, respectively\cite{Lu}, wheras 
the corresponding peaks in Raman scattering are  
at 50 and 60 meV\cite{Limonov}. Excitonic effects in 
the $B_{1g}$
channel thus seem not to be in conflict with available data. The overall 
decrease in intensity as a function of doping of the theoretical B-spectra 
is mainly caused by 
the $\delta^2$ factor in the Raman vertices. It agrees with experiment, 
however, the experimental $B_{1g}$ spectrum decreases faster than the 
$B_{2g}$ spectrum in contrast with the theoretical curves.

 The dotted lines in Fig. 4, called $A_{1g,s}$, show screened
$A_{1g}$ spectra. 
Their absolute intensities are about 2 orders in magnitude smaller than
those for $B_{1g}$ scattering. Moreover, they show in general two
gap features, a weaker one related to the BCS and a stronger one
reflecting the total gap. All these features clearly indicate that
the $A_{1g,s}$ curves are unable to explain the experimental $A_{1g}$ spectra.

According to Ref.\cite{Littlewood} density fluctuations may couple to the 
superconducing order parameter via the modulation of the density of states
at the Fermi energy. This indirect coupling also applies to the
density fluctuations induced by the the $A_{1g}$ Raman vertex $\gamma_1$.
Using the BCS-assumption with a cutoff $\omega_0$ the effective Raman
coupling is 
$\sum_{\bf k}\tilde{\gamma}_1({\bf k}){\tilde c}^\dagger_{{\bf k}\uparrow}
{\tilde c}^\dagger_{{\bf -k}\downarrow}$  + h.c., with $\tilde{\gamma}_1
=g\gamma({\bf k})$ and  
$g=\Delta ln(\omega_0/\Delta)\delta N(0)/N(0)$, where $\delta N(0)$ is the
change in the projected d-wave density at the Fermi surface due to the original
Raman vertex $\gamma_1$. The effective vertex $\tilde{\gamma}_1$ is
unscreened. Moreover, $g$ is large in the optimal and overdoped region because
of the proximity of the van Hove singularity which enhances $\delta N(0)$
and decays rapidly in the underdoped region because of the factor
$\Delta$. The resulting $A_{1g}$ spectrum is again given by the 
diagram in Fig. 2 with external Raman vertices $\tilde{\gamma}_1$ and 
the Heisenberg interaction for the dashed line. Denoting the resulting
susceptibilities with a tilde we obtain
$\tilde{\chi}_1(z) = {\tilde{\chi}^{(0)}_1(z)}/
(1+ J\tilde{\chi}^{(0)}_1(z)/g^2)$,
\begin{eqnarray}
\tilde{\chi}^{(0)}_1(z) = {g^2\over N_c} \sum_{\bf k} \gamma^2({\bf k})
(\Pi_{12,12}({\bf k},z)+\Pi_{11,22}({\bf k},z) \nonumber \\ 
-\Pi_{14,32}({\bf k},z) -\Pi_{13,42}({\bf k},z) +(z \rightarrow -z)).
\label{chi00}
\end{eqnarray}

The dashed and solid lines in Fig. 4, denoted by $\tilde{A}_{1g}^{(0)}$
and $\tilde{A}_{1g}$, show the negative imaginary part of
$\tilde{\chi}^{(0)}_1$ and, $\tilde{\chi}_1$, respectively, for the same
three dopings as in Fig. 3, calculated with the cutoff $\omega_0=J$.
 $\tilde{A}_{1g}^{(0)}$ exhibits a steplike
feature at $2\Delta$ in the overdoped regime which transforms into a
well-pronounced peak towards lower dopings. The peak coincides with the
maximum of the total gap and there is no indication of any contribution 
from the smaller BCS gap, similar as in the $B^{(0)}_{1g}$ spectrum.
Including also vertex corrections spectral weight around $2\Delta$
accumulates into a pronounced collective peak at $2\Delta$ in the overdoped
region describing amplitude fluctuations of the superconducting order 
parameter\cite{Littlewood}. Going from the overdoped to the underdoped regime
this peak first increases, then passes through a maximum around optimal
doping, and then decreases in frequency and intensity, becoming at the same
time rather broad. The decrease 
in its frequency can easily be understood: The denominator of 
$\tilde{\chi}_1(z)$ becomes zero at zero frequency in the limit 
$\Delta \rightarrow 0$ if $\Delta$ satisfies the gap equation.
This implies that the frequency of the collective peak in 
$\tilde{\chi}_1(z)$
has to go to zero at the onset of superconductivity in the underdoped region
in the absence of damping. The density of states is, however, nonzero in
the d-CDW gap even at low frequency. As a result, the collective peak
varies as $2\Delta$ at larger energies but becomes rather broad at low 
energies. This agrees with experiments
in Bi2212 where the $A_{1g}$ peak passes through a maximum
and then decreases on the underdoped side\cite{Kendziora}. 
Taking also data on phonon renormalizations into account
it even has been conjectured\cite{Strohm} in the case of
$YBa_2Cu_3O_7$ that the $A_{1g}$ peak corresponds to the superconducting gap
$2\Delta$.
\vspace{0.3cm} 
\begin{figure}[h]
      \centerline{
      \epsfysize=10cm
      \epsfxsize=10cm
      \epsffile{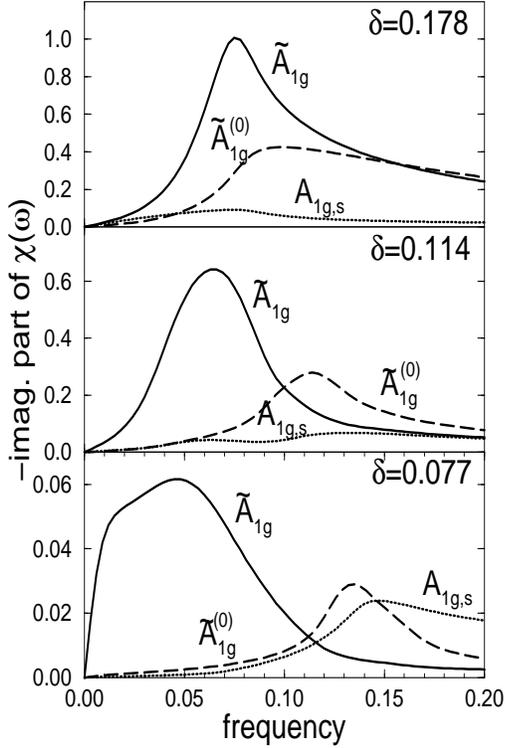}}
\caption 
{$\tilde{A}_{1g}^{(0)}$ and $\tilde{A}_{1g}$: Unperturbed and full 
$A_{1g}$ spectra \\
calculated with the indirect coupling; $A_{1g,s}$:
Screened, full \\
$A_{1g}$ spectrum multiplied
by 50 (upper two diagrams) or 10 \\
(lower digram).}
\label{fig4}
\end{figure}

Fig. 4 shows that $\tilde{A}_{1g}$ is
much larger in magnitude than $A_{1g,s}$ for all dopings,
in spite of the fact that the coupling to the superconducting order
parameter occurs in an indirect way. One reason is that the $\tilde{\gamma}_1$
vertex is unscreened in contrast to $\gamma_1$ and that screening
reduces the intensity by one order
of magnitude or more. A second reason is that the proximity to the van Hove
singularity leads to rather large values for $\delta N(0)/N(0)$ between 2 and 5
in the considered doping region. Comparing the theoretical intensities 
for $B_{1g}$ and $\tilde{A}_{1g}$ scattering with the experimental ones
one finds that $\tilde{A}_{1g}$ has the right magnitude whereas $A_{1g,s}$
would be much too small. We would like to point out, however, that the coupling
constant $g$ also depends on the cutoff $\omega_0$ and thus is subject
to considerable uncertainty.

In conclusion, we have shown that the observed different behavior of the 
three symmetry components of the electronic Raman spectrum in high-T$_c$
superconductors as a function of
doping can be explained within a t-J model in the larg N
limit. Basic ingredients of this approach are the strong competition
of the superconducting and the d-CDW order parameters in the underdoped
regime and the importance of collective effects. The peak in the $B_{1g}$
spectrum in the superconducting state is explained by amplitude fluctuations
of the d-CDW order parameter which, in the optimal and overdoped region, can
also be viewed as excitonic states. We also found that the indirect coupling 
of light to the superconducting order parameter is important leading to the
conclusion that the $A_{1g}$ peak is caused by amplitude fluctuations of the 
superconducting order parameter. 

The authors thank Secyt and the BMBF ( Project ARG 99/007) for financial 
support and C. Bernhard for a critical reading of the manuscript.

\end{multicols}

\begin{thebibliography}{99}

\bibitem{Kendziora} C. Kendziora and A. Rosenberg, Phys. Rev. B {\bf 52},
R9867 (1995)

\bibitem{Strohm} T. Strohm and M. Cardona, Phys. Rev. B {\bf 55}, 12725 (1997)

\bibitem{Opel} M. Opel, R. Nemetschek, C. Hoffmann, R. Philipp, P.F. 
M\"uller, and R. Hackl, Phys. Rev. B {\bf 61}, 9752 (2000)

\bibitem{Venturini} F. Venturini, M. Opel, R. Hackl, H. Berger, L. Forro,
and B. Revaz, cond-mat/0110439

\bibitem{Norman} M.R. Norman, H. Ding, J.C. Campuzano, T. Takeuchi,
M. Randeria, T. Yokoya, T. Takahashi, T. Mochiku, and K. Kadowaki,
Phys. Rev. Lett. {\bf 79}, 3506 (1997)

\bibitem{Cappelluti} E. Cappelluti and R. Zeyher, Phys. Rev. B {\bf 59},
6475 (1999)

\bibitem{Chakravarty} S. Chakravarty, R.B. Laughlin, D.K. Morr, and Ch. Nayak,
Phys. Rev. B {\bf 63}, 94503 (2001)

\bibitem{Zeyher} R. Zeyher and A. Greco, Eur. Phys. J B {\bf 6}, 473 (1998)

\bibitem{Chubukov} A.V. Chubukov, D.K. Morr, and G. Blumberg, Sol. State Comm.
{\bf 112}, 183 (1999)

\bibitem{Devereaux} T.P. Devereaux and A.P. Kampf, Phys. Rev. B {\bf 61},
1490 (2000)

\bibitem{Lu} D.H. Lu, D.L. Feng, N.P. Armitage, K.M. Shen, A. Damascelli,
C. Kim, F. Ronning, Z.-X. Shen, D.A. Bonn, R. Liang, W.N. Hardy,
A.I. Rykov, and S. Tajima, Phys. Rev. Lett. {\bf 86}, 4370 (2001)

\bibitem{Limonov} M.F. Limonov, A.I. Rykov, S. Tajima, and A. Yamanaka,
Phys. Rev. B {\bf 61}, 12412 (2000)

\bibitem{Littlewood} P.B. Littlewood and C.M. Varma, 
Phys. Rev. Lett. {\bf 47}, 811 (1981)




\end{thebibliography}
\end{document}